
%
%
%
%
\def\unredoffs{\hoffset-.14truein\voffset-.2truein} 
 
%
%
\newbox\leftpage \newdimen\fullhsize \newdimen\hstitle \newdimen\hsbody
\tolerance=1000\hfuzz=2pt
\catcode`\@=11 
%
\magnification=1095\unredoffs\baselineskip=16pt plus 2pt minus 1pt
\hsbody=\hsize \hstitle=\hsize 
%
%
%
\newcount\yearltd\yearltd=\year\advance\yearltd by -1900

%
%

\def\draftmode{\message{ DRAFTMODE }\def\draftdate{{\rm preliminary draft:
\number\month/\number\day/\number\yearltd\ \ \hourmin}}%
\headline={\hfil\draftdate}\writelabels\baselineskip=16pt plus 2pt minus 2pt
 {\count255=\time\divide\count255 by 60 \xdef\hourmin{\number\count255}
  \multiply\count255 by-60\advance\count255 by\time
  \xdef\hourmin{\hourmin:\ifnum\count255<10 0\fi\the\count255}}}
\def\nolabels{\def\wrlabeL##1{}\def\eqlabeL##1{}\def\reflabeL##1{}}
\def\writelabels{\def\wrlabeL##1{\leavevmode\vadjust{\rlap{\smash%
{\line{{\escapechar=` \hfill\rlap{\sevenrm\hskip.03in\string##1}}}}}}}%
\def\eqlabeL##1{{\escapechar-1\rlap{\sevenrm\hskip.05in\string##1}}}%
\def\reflabeL##1{\noexpand\llap{\noexpand\sevenrm\string\string\string##1}}}
\nolabels
%
\global\newcount\secno \global\secno=0
\global\newcount\meqno \global\meqno=1
\def\newsec#1{\global\advance\secno by1\message{(\the\secno. #1)}
\global\subsecno=0\eqnres@t\noindent{\bf\the\secno. #1}
\writetoca{{\secsym} {#1}}\par\nobreak\medskip\nobreak}
\def\eqnres@t{\xdef\secsym{\the\secno.}\global\meqno=1\bigbreak\bigskip}
\def\sequentialequations{\def\eqnres@t{\bigbreak}}\xdef\secsym{}
\global\newcount\subsecno \global\subsecno=0
\def\subsec#1{\global\advance\subsecno by1\message{(\secsym\the\subsecno. #1)}
\ifnum\lastpenalty>9000\else\bigbreak\fi
\noindent{\bf\secsym\the\subsecno. #1}\writetoca{\string\quad 
{\secsym\the\subsecno.} {#1}}\par\nobreak\medskip\nobreak}
\def\appendix#1#2{\global\meqno=1\global\subsecno=0\xdef\secsym{\hbox{#1.}}
\bigbreak\bigskip\noindent{\bf Appendix #1. #2}\message{(#1. #2)}
\writetoca{Appendix {#1.} {#2}}\par\nobreak\medskip\nobreak}
%
%
\def\eqnn#1{\xdef #1{(\secsym\the\meqno)}\writedef{#1\leftbracket#1}%
\global\advance\meqno by1\wrlabeL#1}
\def\eqna#1{\xdef #1##1{\hbox{$(\secsym\the\meqno##1)$}}
\writedef{#1\numbersign1\leftbracket#1{\numbersign1}}%
\global\advance\meqno by1\wrlabeL{#1$\{\}$}}
\def\eqn#1#2{\xdef #1{(\secsym\the\meqno)}\writedef{#1\leftbracket#1}%
\global\advance\meqno by1$$#2\eqno#1\eqlabeL#1$$}
%
\newskip\footskip\footskip14pt plus 1pt minus 1pt 
\def\footnotefont{\ninepoint}\def\f@t#1{\footnotefont #1\@foot}
\def\f@@t{\baselineskip\footskip\bgroup\footnotefont\aftergroup\@foot\let\next}
\setbox\strutbox=\hbox{\vrule height9.5pt depth4.5pt width0pt}
\global\newcount\ftno \global\ftno=0
\def\foot{\global\advance\ftno by1\footnote{$^{\the\ftno}$}}
%
\newwrite\ftfile   
\def\footend{\def\foot{\global\advance\ftno by1\chardef\wfile=\ftfile
$^{\the\ftno}$\ifnum\ftno=1\immediate\openout\ftfile=foots.tmp\fi%
\immediate\write\ftfile{\noexpand\smallskip%
\noexpand\item{f\the\ftno:\ }\pctsign}\findarg}%
\def\footatend{\vfill\eject\immediate\closeout\ftfile{\parindent=20pt
\centerline{\bf Footnotes}\nobreak\bigskip\input foots.tmp }}}
\def\footatend{}
%
%
\global\newcount\refno \global\refno=1
\newwrite\rfile
\def\ref{[\the\refno]\nref}
\def\nref#1{\xdef#1{[\the\refno]}\writedef{#1\leftbracket#1}%
\ifnum\refno=1\immediate\openout\rfile=refs.tmp\fi
\global\advance\refno by1\chardef\wfile=\rfile\immediate
\write\rfile{\noexpand\item{#1\ }\reflabeL{#1\hskip.31in}\pctsign}\findarg}
\def\findarg#1#{\begingroup\obeylines\newlinechar=`\^^M\pass@rg}
{\obeylines\gdef\pass@rg#1{\writ@line\relax #1^^M\hbox{}^^M}%
\gdef\writ@line#1^^M{\expandafter\toks0\expandafter{\striprel@x #1}%
\edef\next{\the\toks0}\ifx\next\em@rk\let\next=\endgroup\else\ifx\next\empty%
\else\immediate\write\wfile{\the\toks0}\fi\let\next=\writ@line\fi\next\relax}}
\def\striprel@x#1{} \def\em@rk{\hbox{}} 
\def\lref{\begingroup\obeylines\lr@f}
\def\lr@f#1#2{\gdef#1{\ref#1{#2}}\endgroup\unskip}
\def\semi{;\hfil\break}
\def\addref#1{\immediate\write\rfile{\noexpand\item{}#1}} 
\def\startrefs#1{\immediate\openout\rfile=refs.tmp\refno=#1}
\def\xref{\expandafter\xr@f}\def\xr@f[#1]{#1}
\def\refs#1{\count255=1[\r@fs #1{\hbox{}}]}
\def\r@fs#1{\ifx\und@fined#1\message{reflabel \string#1 is undefined.}%
\nref#1{need to supply reference \string#1.}\fi%
\vphantom{\hphantom{#1}}\edef\next{#1}\ifx\next\em@rk\def\next{}%
\else\ifx\next#1\ifodd\count255\relax\xref#1\count255=0\fi%
\else#1\count255=1\fi\let\next=\r@fs\fi\next}
%

%
\newwrite\ffile\global\newcount\figno \global\figno=1
\def\fig{fig.~\the\figno\nfig}
\def\nfig#1{\xdef#1{fig.~\the\figno}%
\writedef{#1\leftbracket fig.\noexpand~\the\figno}%
\ifnum\figno=1\immediate\openout\ffile=figs.tmp\fi\chardef\wfile=\ffile%
\immediate\write\ffile{\noexpand\medskip\noexpand\item{Fig.\ \the\figno. }
\reflabeL{#1\hskip.55in}\pctsign}\global\advance\figno by1\findarg}
\def\xfig{\expandafter\xf@g}\def\xf@g fig.\penalty\@M\ {}
\def\figs#1{figs.~\f@gs #1{\hbox{}}}
\def\f@gs#1{\edef\next{#1}\ifx\next\em@rk\def\next{}\else
\ifx\next#1\xfig #1\else#1\fi\let\next=\f@gs\fi\next}
\newwrite\lfile
{\escapechar-1\xdef\pctsign{\string\%}\xdef\leftbracket{\string\{}
\xdef\rightbracket{\string\}}\xdef\numbersign{\string\#}}

\def\writestop{\def\writestoppt{\immediate\write\lfile{\string\pageno%
\the\pageno\string\startrefs\leftbracket\the\refno\rightbracket%
\string\def\string\secsym\leftbracket\secsym\rightbracket%
\string\secno\the\secno\string\meqno\the\meqno}\immediate\closeout\lfile}}
\def\writestoppt{}\def\writedef#1{}
\def\seclab#1{\xdef #1{\the\secno}\writedef{#1\leftbracket#1}\wrlabeL{#1=#1}}
\def\subseclab#1{\xdef #1{\secsym\the\subsecno}%
\writedef{#1\leftbracket#1}\wrlabeL{#1=#1}}
\newwrite\tfile \def\writetoca#1{}
\def\leaderfill{\leaders\hbox to 1em{\hss.\hss}\hfill}
\def\writetoc{\immediate\openout\tfile=toc.tmp 
   \def\writetoca##1{{\edef\next{\write\tfile{\noindent ##1 
   \string\leaderfill {\noexpand\number\pageno} \par}}\next}}}

%
\catcode`\@=12 
%
\edef\tfontsize{\ifx\answ\bigans scaled\magstep3\else scaled\magstep4\fi}
 \tfontsize  \tfontsize
 \tfontsize \font\titlei=cmmi10 \tfontsize
\font\titleis=cmmi7 \tfontsize \font\titleiss=cmmi5 \tfontsize
\font\titlesy=cmsy10 \tfontsize \font\titlesys=cmsy7 \tfontsize
\font\titlesyss=cmsy5 \tfontsize  \tfontsize
\skewchar\titlei='177 \skewchar\titleis='177 \skewchar\titleiss='177
\skewchar\titlesy='60 \skewchar\titlesys='60 \skewchar\titlesyss='60
 \ifx\answ\bigans\else scaled\magstep1\fi
\ifx\answ\bigans\else

 \font\absi=cmmi10 scaled\magstep1
\font\absis=cmmi7 scaled\magstep1 \font\absiss=cmmi5 scaled\magstep1
\font\abssy=cmsy10 scaled\magstep1 \font\abssys=cmsy7 scaled\magstep1
\font\abssyss=cmsy5 scaled\magstep1 
\skewchar\absi='177 \skewchar\absis='177 \skewchar\absiss='177
\skewchar\abssy='60 \skewchar\abssys='60 \skewchar\abssyss='60
\fi
\font\ninerm=cmr9 \font\sixrm=cmr6 \font\ninei=cmmi9 \font\sixi=cmmi6 
\font\ninesy=cmsy9 \font\sixsy=cmsy6 \font\ninebf=cmbx9 
\font\nineit=cmti9 \font\ninesl=cmsl9 \skewchar\ninei='177
\skewchar\sixi='177 \skewchar\ninesy='60 \skewchar\sixsy='60 
\def\ninepoint{\def\rm{\fam0\ninerm}
\textfont0=\ninerm \scriptfont0=\sixrm \scriptscriptfont0=\fiverm
\textfont1=\ninei \scriptfont1=\sixi \scriptscriptfont1=\fivei
\textfont2=\ninesy \scriptfont2=\sixsy \scriptscriptfont2=\fivesy
\textfont\itfam=\ninei \def\it{\fam\itfam\nineit}\def\sl{\fam\slfam\ninesl}%
\textfont\bffam=\ninebf \def\bf{\fam\bffam\ninebf}\rm} 
%
%

\hyphenation{anom-aly anom-alies coun-ter-term coun-ter-terms}
\def\inv{^{\raise.15ex\hbox{${\scriptscriptstyle -}$}\kern-.05em 1}}

\def\Dsl{\,\raise.15ex\hbox{/}\mkern-13.5mu D} 
\def\dsl{\raise.15ex\hbox{/}\kern-.57em\partial}

\def\lspace{\ifx\answ\bigans{}\else\qquad\fi}
\def\lbspace{\ifx\answ\bigans{}\else\hskip-.2in\fi} 
\def\boxeqn#1{\vcenter{\vbox{\hrule\hbox{\vrule\kern3pt\vbox{\kern3pt
	\hbox{${\displaystyle #1}$}\kern3pt}\kern3pt\vrule}\hrule}}}
\def\mbox#1#2{\vcenter{\hrule \hbox{\vrule height#2in
		\kern#1in \vrule} \hrule}}  
%

\def\darr#1{\raise1.5ex\hbox{$\leftrightarrow$}\mkern-16.5mu #1}

\def\half{{\textstyle{1\over2}}} 
\def\roughly#1{\raise.3ex\hbox{$#1$\kern-.75em\lower1ex\hbox{$\sim$}}}

\def\p2inf{\mathrel{\mathop{\sim}\limits_{\scriptscriptstyle
{p^2 \rightarrow \infty }}}}
\def\kap2inf{\mathrel{\mathop{\sim}\limits_{\scriptscriptstyle
{\kappa \rightarrow \infty }}}}
\def\x2inf{\mathrel{\mathop{\sim}\limits_{\scriptscriptstyle
{x \rightarrow \infty }}}}
\def\Lam2inf{\mathrel{\mathop{\sim}\limits_{\scriptscriptstyle
{\Lambda \rightarrow \infty }}}}
\def\frac#1#2{{{#1}\over {#2}}}
\def\half{\hbox{${1\over 2}$}}

\def\Mev{{\rm MeV}}\def\Gev{{\rm GeV}}

\def\lsim{\mathrel{mathpalette\@v1000ersim<}}
\def\gsim{\mathrel{mathpalette\@versim>}}

\catcode`@=11 
\def\slash#1{\mathord{\mathpalette\c@ncel#1}}
 \def\c@ncel#1#2{\ooalign{$\hfil#1\mkern1mu/\hfil$\crcr$#1#2$}}
\def\lsim{\mathrel{\mathpalette\@versim<}}
\def\gsim{\mathrel{\mathpalette\@versim>}}
 \def\@versim#1#2{\lower0.2ex\vbox{\baselineskip\z@skip\lineskip\z@skip
       \lineskiplimit\z@\ialign{$\m@th#1\hfil##$\crcr#2\crcr\sim\crcr}}}
\catcode`@=12 

\def\PR{{\it Phys.~Rev.~}}
\def\PRL{{\it Phys.~Rev.~Lett.~}}
\def\NP{{\it Nucl.~Phys.~}}
\def\PL{{\it Phys.~Lett.~}}

\def\ZP{{\it Zeit.~Phys.~}}

\def\vol#1{{\bf #1}}
\def\vyp#1#2#3{\vol{#1} (#2) #3}

\def\Asl{\raise.15ex\hbox{/}\mkern-11.5mu A}
\def\psl{\lower.12ex\hbox{/}\mkern-9.5mu p}
\def\qsl{\lower.12ex\hbox{/}\mkern-9.5mu q}
\def\rsl{\lower.03ex\hbox{/}\mkern-9.5mu r}
\def\ksl{\raise.06ex\hbox{/}\mkern-9.5mu k}
\def\Mev{\hbox{MeV}}


\pageno=0\nopagenumbers\tolerance=10000\hfuzz=5pt
\line{\hfill RAL-TR-97-039}
\vskip 36pt
\centerline{\bf What Have We Learned About and From $F_L(x,Q^2)$ at HERA?}
\vskip 36pt
\centerline{Robert~S.~Thorne}
\vskip 12pt
\centerline{\it Rutherford Appleton Laboratory,}
\centerline{\it Chilton, Didcot, Oxon., OX11 0QX, U.K.}
\vskip 0.9in
{\narrower\baselineskip 10pt
\centerline{\bf Abstract}
\medskip
Recently  the H1 collaboration has published a ``determination'' of the 
structure function $F_L(x,Q^2)$ at low $x$. I address the question of how 
reliable this determination really is. I argue that it is in fact  
a consistency check of a given theoretical approach 
rather than a real determination
of the value of $F_L(x,Q^2)$, but potentially a very useful one.
I compare the consistency of different approaches, and indeed find that a
LO--in--$\alpha_s$ calculation of structure functions is completely ruled out.
I also find that the ``determined'' values of $F_L(x,Q^2)$ are 
surprisingly stable under changes in theoretical approach but, when 
working consistently within a well-defined theoretical 
framework, the values of $F_L(x,Q^2)$ implied are somewhat lower than 
previously quoted.}
   
\vskip 0.7in
\line{RAL-TR-97-039\hfill}
\line{August 1997\hfill}
\vfill\eject
\footline={\hss\tenrm\folio\hss}



\newsec{Introduction}

The recent measurements of $F_2(x,Q^2)$ at HERA 
\ref\hone{H1 collaboration,
\NP B \vyp{470}{1996}{3}.}\ref\zeus{ZEUS collaboration: M. Derrick {\it et al},
\ZP C \vyp{69}{1996}{607}\semi Preprint DESY 96-076 (1996), to be published
in \ZP C.} have provided data on
a structure function at far lower values of $x$ than any previous
experiments. However, these measurements are not of the structure function 
directly. Rather they are of the differential cross--section 
\eqn\one{{d^2 \sigma\over dxdQ^2}\equiv {2\pi \alpha^2 \over xQ^4}
\biggl[(2(1-y)+y^2)F_2(x,Q^2) -y^2F_L(x,Q^2)\biggr].}
$Q^2$ is the squared four--momentum transferred in the lepton--proton
scattering, $x$ is the scaling variable and $y=Q^2/sx$ where $s$ is the centre
of mass energy squared. At HERA $s=90200\Gev^2$. However, $F_L(x,Q^2)$ is 
expected to be much smaller than $F_2(x,Q^2)$ for most
of the parameter space in which measurements take place
(it must be smaller than $F_2(x,Q^2)$ at all $x$ and $Q^2$), and within this 
parameter space $y$ is nearly always small ($0.25$ or less). Hence, the 
measurement is usually effectively of $F_2(x,Q^2)$. 

However, at the fringe of parameter space, i.e. the highest $Q^2$ values for
the lowest $x$ measurements, the values of $y$ can be somewhat larger, 
reaching $\approx 0.7$ at their maximum. For values of $y$ larger
than about $0.25$ the fact that $F_L(x,Q^2)$ is exprected to have a 
non--zero value starts affecting the value of $F_2(x,Q^2)$ extracted 
from the measurement of 
the differential cross--section by about $1\%$. This increases to about 
$10-20\%$ at the highest $y$ values of $0.7$ 
(clearly depending on the value of $F_L(x,Q^2)$ used).
Hence in this region of parameter space, within the typical quoted errors,
the HERA collaborations cannot be claiming to measure $F_2(x,Q^2)$
directly.
It would perhaps be simplest to release data in this region of parameter space 
in the form of the differential cross--section rather than in terms of 
structure functions. However, both H1 and ZEUS choose to take the values of 
$F_L(x,Q^2)$ from some theoretical prediction, and extract the
consequent ``measured'' values of $F_2(x,Q^2)$. 
Both collaborations obtain their
values of $F_L(x,Q^2)$ from fits to the data using the usual
approach of solving renormalization group equations for parton 
distributions and combining with coefficient functions for 
parton scattering with the calculations done to NLO--in-$\alpha_s$.   
In \hone\ $F_L(x,Q^2)$ is calculated using the 
GRV parameterization of partons \ref\grv{M. Gl\"uck, E. Reya and A. Vogt,
\ZP \vyp{C67}{1995}{433}.}, while in both of \zeus\ it is 
calculated using the 
parton distributions obtained from their own iterative fit to $F_2(x,Q^2)$. 

Thus, both collaborations produce values of $F_2(x,Q^2)$ which 
take into account the values of $F_L(x,Q^2)$ in an approximate
manner. However, it is 
very important to remember that the values of $F_2(x,Q^2)$ quoted in \hone\
and \zeus\ are not really measurements of $F_2(x,Q^2)$ in the high $y$ region,
and any attempt to 
fit the HERA data using some theoretical model should take this into account.
The values of $F_2(x,Q^2)$ used in a fit should really be those obtained
from the measured differential cross--section and the prediction for 
$F_L(x,Q^2)$ consistent with the particular theoretical approach (it is
difficult to imagine a genuine theoretical approach which does not 
calculate $F_L(x,Q^2)$ alongside $F_2(x,Q^2)$). Alternatively, the fit 
could be directly to the cross--section using the calculated values of both 
$F_2(x,Q^2)$ and $F_L(x,Q^2)$. (The two approaches are identical if the 
error on $F_2(x,Q^2)$ in the former does not change with the 
predicted $F_L(x,Q^2)$, but is equal to its value when $F_L(x,Q^2)=0$.)
For any ``conventional''
NLO--in--$\alpha_s$  approach the differences will not be all that large,
but when using different approaches, such as 
those involving summation of leading logs in $(1/x)$, or even a 
LO--in--$\alpha_s$ calculation, the differences in the predictions for 
$F_L(x,Q^2)$ can be large enough that the extracted values of $F_2(x,Q^2)$ 
move by amounts greater than their errors on the fringes of parameter 
space. Moreover, the direction of movement tends to be the same for every 
point, so ignoring this effect can have a sizeable effect on the fit. 
Unfortunately the 
consistent manner of comparing theory to $F_2(x,Q^2)$ 
values described above seems not to 
be done in practically all fits to data. 

Recently the H1 collaboration took a rather different approach to the 
treatment of their high $y$ data \ref\flong{H1 Collaboration, \PL 
\vyp{B393}{1997}{452}.}.
Rather than adopt the procedure outlined above, 
they fitted the data for $F_2(x,Q^2)$ in the 
region of low $y$ using NLO--in--$\alpha_s$ QCD, used this fit to extrapolate 
$F_2(x,Q^2)$ into the high $y$ region, and from the difference between
this extrapolation and the measured cross--section obtained a ``measurement''
of $F_L(x,Q^2)$. This determination of the value of $F_L(x,Q^2)$
relies only on assuming the 
correctness of the NLO--in--$\alpha_s$ fit for $F_2(x,Q^2)$. But if one is 
to assume this correctness then it seems perverse indeed not to assume the
correctness of the NLO--in--$\alpha_s$ $F_L(x,Q^2)$ that is predicted as a 
result. Hence, {\it if the theoretical
approach is correct}, $F_L(x,Q^2)$ is already determined.
The difference between the measured cross--section and
the extrapolated $F_2(x,Q^2)$ then provides nothing more than a 
consistency check: if it disagrees with the predicted $F_L(x,Q^2)$ it 
suggests that the theory is wrong, if it agrees then the theory is not 
necessarily wrong. This is all that one can conclude. There may well be
other theoretical approaches which fit the low $y$ data equally well
(or better) but which have different extrapolations and/or predictions for
$F_L(x,Q^2)$. Indeed, in the small $x$ region this is 
likely since higher orders in $\alpha_s$ are accompanied by higher 
order terms in $\ln(1/x)$, and these leading $\ln(1/x)$ terms may well
introduce important corrections to the standard approach. 
If the extrapolations and predictions match for these other
approaches, then the predicted $F_L(x,Q^2)$ is just as likely to be correct,
but is no more a real measurement than any other matching case. 

Hence, the procedure adopted by the H1 collaboration is not in itself a
real measurement of $F_L(x,Q^2)$. However, it is potentially a very 
useful way of discriminating between different theoretical approaches. 
The consistency of
the predicted $F_L(x,Q^2)$ and the ``measured'' $F_L(x,Q^2)$ for a given 
approach is a very non--trivial check on the theory since, if the free 
parameters in the theory are already tied down by a fit for high $y$ data, 
then it is a parameter free check on a particular relationship between 
$F_2(x,Q^2)$ and $F_L(x,Q^2)$. Hence, in this paper I will present the
approach used by H1 in the manner of a consistency check. 
I will do this for four 
different types of approach: a fit analogous to that of H1 in \flong,
which I will demonstrate is not actually a useful NLO--in $\alpha_s$ fit: 
a more correct NLO--in--$\alpha_s$ fit: a LO--in--$\alpha_s$ fit: 
and a fit using 
the LORSC approach which includes leading $\ln(1/x)$ terms\ref\mypaper{R.S. 
Thorne, \PL \vyp{B392}{1997}{463}: {\tt hep-ph/9701241}.}. I will also include
more data at high $y$ than that in $\flong$.
I will find that the ``measured'' $F_L(x,Q^2)$  is very similar in 
each of the three latter approaches, but lower than in the first one.
However, this similarity would start to disappear at higher $y$. 
I also find that find that the data with its current errors 
is not yet sufficient to rule out any approach other than the 
LO--in--$\alpha_s$ fit, which it does rule out very clearly.
One can, if one wants, interpret the 
``measurements'' of $F_L(x,Q^2)$  within the two approaches satisfying
consistency to be the likely values of $F_L(x,Q^2)$, 
but this is not really fundamentally different from quoting instead the values 
of $F_L(x,Q^2)$ predicted within either of these approaches.  

\newsec{Comparison of Different Theoretical Approaches.}

I compare the four different approaches outlined above. First I
discuss the data to be used and the accuracy of $F_L(x,Q^2)$ achieved. 
I define a rescaled differential cross--section
\eqn\two{\tilde\sigma(x,Q^2)= {Q^4x\over 2\pi \alpha^2}{1\over (2(1-y)+y^2)}
{d^2\sigma\over dxdQ^2}.}
Hence I may write
\eqn\three{\tilde\sigma(x,Q^2)=F_2(x,Q^2)-Y^{-1}\cdot F_L(x,Q^2),\hskip 0.5in 
Y={(2(1-y)+y^2) \over y^2  }.}
This is clearly a useful definition since in the limit 
$F_L(x,Q^2)\to 0$ or  $y\to 0$, $\tilde \sigma
=F_2(x,Q^2)$. Hence it allows a simple parameterization of the effect 
of non--zero $F_L(x,Q^2)$ on the extracted $F_2(x,Q^2)$. 
From \three\ it is clear that denoting $F_2(x,Q^2)$ obtained from the 
the fit by $F_2^p(x,Q^2)$, the ``measured'' value of 
$F_L(x,Q^2)$, denoted by $F^m_L(x,Q^2)$, is given by  
\eqn\threea{F^m_L(x,Q^2) =Y(F^p_2(x,Q^2)-\tilde\sigma (x,Q^2)),}
and that the error in $F^m_L(x,Q^2)$ is
\eqn\four{\Delta F^m_L(x,Q^2)=Y\cdot\Bigl[(\Delta \tilde\sigma(x,Q^2))^2
+(\Delta F^p_2(x,Q^2))^2\Bigr]^{\half},}
where $\Delta \tilde \sigma(x,Q^2) >> \Delta F^p_2(x,Q^2)$ in practice. Hence, 
the error on $F^m_L(x,Q^2)$ depends linearly on the 
error on the measurement of the cross--section and roughly quadratically
on $y^{-1}$. Although this means that, if one wishes to make a sensible 
consistency check for $F^m_L(x,Q^2)$ 
(by comparing to the predicted $F_L^p(x,Q^2)$), 
it is necessary to have high $y$, a relatively small
decrease in $y$ can be countered by a large decrease in the error on the 
measurement of $\tilde \sigma(x,Q^2)$. For example, an error of $5\%$
on the measurement of $\tilde \sigma (x,Q^2)$  
at $y=0.6$ leads to a more accurate determination of $F_L^m(x,Q^2)$ than 
an error of $10\%$ on the measurement of $\tilde \sigma (x,Q^2)$ at $y=0.7$. 
Examining the full range of 
data in \hone\ and \zeus\ there are a number of points which give an accuracy
of measurement of $F^m_L(x,Q^2)$ which is comparable to the 6 points in 
\flong. Taking a cut on the error produced for $F^m_L(x,Q^2)$ to be $0.3$
and a lower cut on $Q^2$ of $5\Gev^2$ (in order not to be too close to the 
charm threshold), I find that
there is one point in \hone, and 8 points in the latter of \zeus.
Of these additional points, 5 of the ZEUS points have $y$ comparable to the
$0.7$ in \flong, while the other 3 ZEUS points and the H1 point have somewhat 
smaller $y$ but better accuracy in measurement. The full set of data points 
to be used is shown in table 1. 

Now that the relevant data are defined, let me consider the fits. First I 
note that the heavy quark thresholds are treated rather differently than in 
\flong. Rather than using the NLO--in--$\alpha_s$ fixed flavour scheme, where 
the charm contribution to the structure function is entirely generated by
coefficient functions convoluted with light parton distributions, 
I simply change the number of active quark 
flavours discontinuously at $m_h^2=Q^2$, and treat the heavy quarks as 
massless above this. I take $m_c^2=2.75\Gev^2$ in order to give a good 
description of the charm data\ref\charmone{EMC collaboration: J.J. Aubert
{\it et al.}, \NP \vyp{B213}{1983}{31}.}\ref\charmtwo{H1 collaboration: 
C. Adloff {\it et al.}, \ZP \vyp{C72}{1996}{593}.}, and choose 
$m_b^2=20\Gev^2$. Neither of the 
above approaches for the treatment of charm is entirely satisfactory since 
the former does not sum leading
logarithms in $Q^2/m_c^2$ while the latter does not treat the threshold 
in a correct manner. In practice the former is more accurate for 
$Q^2<10\Gev^2$ and the latter more accurate for $Q^2>20\Gev^2$. For 
$10\Gev^2<Q^2<20\Gev^2$ the approaches are of roughly equal validity. 

In comparing the predicted 
and measured $F_L(x,Q^2)$ I adopt a different procedure from H1 and 
assume that the variation in $F^p_2(x,Q^2)$ is 
entirely due to variations within a given theoretical approach 
fitting a particular set of data. The difference between the 
various theoretical approaches, and/or from using different sets of data will
be seen in the four different sets of results. 
Within a given approach the variation in $F^p_2(x,Q^2)$
is due only to letting the 
$\chi^2$ for the fit vary to the extent that the quality has a
confidence level only a little lower than the absolute best fit. 
I will discuss the results of this later, but mention here that the errors are 
very small. Also, since I have as much faith in the 
NLO--in--$\alpha_s$ calculation of $F^p_L(x,Q^2)$ as of $F^p_2(x,Q^2)$, 
and hence am only performing a consistency check, I include all
data other than those points in table 1 in the fit, and let the value of
$F^p_L(x,Q^2)$ used in the extraction of
$F^m_2(x,Q^2)$ for these points be that predicted by the parameters determined
in the fit (i.e. the fit is iterative). Thus, the procedure differs 
from \flong\ in 
including the small number of points with $y>0.35$ not in table 1,
which has an extremely small effect on the fit, and also by using a particular
$F^p_L(x,Q^2)$ for the points in the fit rather than letting it vary over
the rather extreme range of $0\to F_2(x,Q^2)$. Avoiding this second variation
means that I do not have the large systematic error in the extrapolated 
$F^p_2(x,Q^2)$ which is seen in \flong. Overall the method of defining the 
uncertainties due to the 
fitting assumptions reduces them immensely compared to those in \flong.

\medskip

\noindent 1. First I consider the analogous fit to that performed by H1. In 
this the fit is performed to BCDMS \ref\bcdms{BCDMS collaboration: A.C. 
Benvenuti {\it et al.}, \PL \vyp{B223}{1989}{485}.} and H1 data only. 
This leads to a problem 
often encountered (or rather ignored) by those attempting to fit small $x$ 
structure function data: the parameters defining the parton distributions 
which are determined by the best fit turn out to be completely 
incompatible with some perfectly respectable data which have not been included
in the fit. The parameters determined by the the fit 
are not included in \flong. However, performing a fit to precisely the same 
data I obtained a gluon distribution which is far too small at $x>0.1$ to
be consistent with the WA70 prompt photon data \ref\wa{WA70 collaboration:
M. Bonesini {\it et al.}, \ZP \vyp{C38}{1988}{371}.}, and is even much smaller 
than the gluon produced by the H1 collaboration in the fit in \hone. 
Although there is a certain 
amount of uncertainty about the true accuracy of the prompt photon 
data, the gluon 
obtained is very much smaller than any possible lower limit on the data. 
It is also far too small to be consistent with the EMC charm data \charmone\ 
using any sensible treatment of charm, and I note that the gluon required by 
this charm data is actually similar to that required by the WA70 prompt photon 
data, i.e. $\sim 2.5(1-x)^6$ at $Q^2=5\Gev^2$. 
Since there is so little gluon at high $x$ in my fit, the momentum sum 
rule allows much more at small $x$. This enables the H1 data to be 
fit well even using a low value
of $\Lambda_{QCD}$ which is preferred by the BCDMS data. Indeed, my fit 
chooses $\Lambda_{QCD}^{n_f=4}=186\Mev$ which is very similar to the
H1 $\Lambda_{QCD}^{n_f=3}=210\Mev$. This should be compared with those 
analyses of only H1 data which produce $\Lambda_{QCD}^{n_f=4}\approx 350\Mev$
when the high $x$ parton distributions are constrained differently
\ref\ballfort{R.D. Ball and S. Forte, 
Proc. of the International Workshop on Deep
Inelastic Scattering, Rome, April, 1996, p. 208.}.     

Ignoring this very serious problem, the fit is performed as in 
\flong, and considers only the same points as \flong\ for measuring 
$F_L(x,Q^2)$. As already mentioned, the only uncertainty
in $F^p_2(x,Q^2)$ is that obtained in allowing the quality of the fit to vary.
For this fit to a relatively small number of data points the variation in 
$F_2^p(x,Q^2)$ while maintaining a ``good'' quality fit is not 
insignificant. However, since I am not really taking this fit 
seriously anyway, but are only using it for illustrative purposes, I quote
the error as being due to the cross--section measurement only.
As one would expect, the results of this procedure
are very similar to those in \flong. The fit to the small $x$
H1 data is very similar, as is the extrapolation. The 
values of $F^m_L(x,Q^2)$ for the $6$ data points for 
$\tilde\sigma$ used in \flong\ are shown in table 2 along with the values
``measured'' in \flong (denoted by $F^m_L(x,Q^2)[4]$) and the values of   
of $F^p_L(x,Q^2)$. The results are also displayed in \fig\figone{Result of 
the measured values of $F^m_L(x,Q^2)$ for the six points used in \flong, 
along with the predicted value, $F^p_L(x,Q^2)$, along a curve of constant 
$y=0.7$ (the solid line), when using a NLO--in--$\alpha_s$ fit 
analogous to that in \flong.
Also shown are the measured values of $F_L(x,Q^2)$ in \flong, 
$F^m_L(x,Q^2)[4]$, and the curve of predicted values in this paper 
(dashed curve).}, where they are compared directly with those in \flong.
One can see that the results for $F^m_L(x,Q^2)$ are indeed very similar to 
those in \flong. The predicted $F_L(x,Q^2)$ is very slightly higher 
than in \flong\ at lowish $Q^2$ but the two converge at higher $Q^2$, 
which can be interpreted as the effects of the different treatment of charm  
in the two approaches not compensating exactly at lowish $Q^2$, but
the difference disappearing at higher $Q^2$, as expected.
Hence, I have very good compatibility with \flong\ so far.

I am now able to explore the implications of this type of fit quantitatively. 
I do indeed have the previously mentioned dramatic inconsistency with the 
prompt photon data and the EMC 
charm data. Also, if I simply put the parameters obtained by the 
previous fit into a global fit to structure function data, i.e include 
NMC \ref\nmc{NMC collaboration: M. Arneodo {\it et al.}, \NP 
\vyp{B483}{1997}{3}.}, CCFR \ref\ccfr{CCFR collaboration: 
P.Z. Quintas {\it et al.}, \PRL 
\vyp{71}{1993}{1307}.} and ZEUS data, then I find that the fit to 
ZEUS data is a little worse
than that obtained form a global NLO--in--$\alpha_s$ fit, and the fit to 
NMC and CCFR data is very much worse. Overall, the global fit is 
completely uncompetitive with the fits produced by e.g. MRS or CTEQ even 
ignoring the particular problem of the high $x$ gluon. 
Hence, the type of fit performed 
above certainly does not lead to a correct parameterization of the parton 
distributions and hence extrapolation of the structure function. Thus, even 
assuming the correctness of the NLO--in--$\alpha_s$ calculation of structure 
functions one does not really learn anything concrete about $F_L(x,Q^2)$
from the above fit, or from \flong. 

\medskip

\noindent 2. In order to make a more reliable consistency check for the high 
$y$ data
I now consider the implications of using precisely the same theoretical 
framework but including more data in the fit.
In order to see the correct results of using the  
NLO--in--$\alpha_s$ calculation I must determine the free parameters in the 
calculation by fitting to all available structure function data, with
some appropriate cuts, and also imposing the constraint that the charm
structure function data from EMC \charmone\ and H1 \charmtwo\ 
is also described well. In practice I 
perform the fit to data in precisely the same manner as in 
\mypaper\ except for one minor point. The ZEUS shifted vertex data
in the latter of \zeus\ seems to be systematically larger than the data in the 
former. Hence I allow their normalization to be 0.98 that of the nominal 
vertex data, i.e. the amount allowed by the uncorrelated errors on 
normalization (the fit would actually prefer 0.97). Hence, the fit is 
almost identical to the NLO1 fit in \mypaper, but this shift in normalization 
allows the $\chi^2$ to be a few points better. 
It is qualitatively a great deal 
different from the fit discussed above. The constraint on the gluon at high 
$x$ results in there being much less gluon at small $x$ and the value of 
$\Lambda_{QCD}^{n_f=4}$ rises to $315\Mev$ in order to produce a satisfactory 
fit to small $x$ data. This rise in $\Lambda_{QCD}^{n_f=4}$ results in a 
worse fit to BCDMS data, but the total fit quality is a great deal better
than that for the type of parameterization above. In the fit the 
normalization required by the H1 data is $0.985$ and for the ZEUS data it is 
$0.99$. The quality of the absolute best fit has a rather low confidence level
(assuming Gaussian errors) of $6\%$. Letting this reduce to 
$1\%$ results in a variation of $F_2^p(x,Q^2)$ which when inserted in \four\
leads to negligible effect and I take the error to be that due to 
$\Delta \tilde \sigma (x,Q^2)$ alone.

Using this fit the form of $F_2^p(x,Q^2)$ is rather different at small $x$
from that obtained using the first procedure. 
The growth with $Q^2$ is slower than above, even taking into account the
different normalization of the data, and hence in a given $x$ bin the value
of $F^p_2(x,Q^2)$ becomes progressively smaller as $y$ increases. 
From \threea\ it is 
obvious that this will result in a lower value of $F_L^m(x,Q^2)$. 
The results for $F_L^m(x,Q^2)$, along with $F_L^p(x,Q^2)$
are shown in table 3 and displayed in \fig\figtwo{Result of 
the measured values of $F^m_L(x,Q^2)$ for the 15 points satisfying the 
cut of 0.3 on the error, along with the predicted values, $F^p_L(x,Q^2)$,
for a NLO--in--$\alpha_s$ global fit. 
In the top figure the points all have $y\approx 0.7$ and $F^p_L(x,Q^2)$
is along a curve of constant $y=0.7$ (the solid line). In the lower figure 
the points have $y \approx 0.55$ and $F^p_L(x,Q^2)$ is along a curve of 
constant $y=0.55$.} where now I use all 15 data
points. (I display the extracted points with a negative value of 
$F^m_L(x,Q^2)$ if this is what is obtained from the use of \three,
even though this is forbidden in practice.)
In the top figure all data with $y\approx 0.7$ is shown, while the 
lower figure contains the points with rather lower $y$.
The measured values are indeed smaller than previously.
Compared with the values in 
\flong\ $F_L^m(x,Q^2)$ is about $0.1$ smaller for the lowest $x$ points 
increasing 
to more than $0.15$ lower for the highest $x$ and $Q^2$ points. Hence, for 
some of these points the value of $F_L^m(x,Q^2)$ reduces to less than 
$60\%$ of its value in \flong. By examination of fig. 2 one sees that, as well 
as the values of $F_L^m(x,Q^2)$ for the 6 points corresponding to those in 
\flong\ having reduced significantly, the inclusion of the other data points 
tends also to imply a somewhat lower value of $F_L^m(x,Q^2)$. A simple
unweighted average of the 6 points in \flong\ gives $F_L^A(x,Q^2)=0.48$. 
The same 6 points in this extraction yield $F_L^A(x,Q^2)=0.34$, while for all
11 points with $y\approx 0.7$ I get $F_L^A(x,Q^2)=0.27$.  I claim that 
{\it if one believes the 
validity of the NLO--in--$\alpha_s$ calculation} then the values in table 3
are the most reliable extractions of $F_L^m(x,Q^2)$ since they result from a 
much more realistic fit. However, taking these values results in a rather
different quantitative conclusion on the value of $F_L^m(x,Q^2)$ for 
$y\approx 0.7$ for $0.001>x>0.0001$ from that reached in \flong, i.e. 
it is $\approx 60\%$ of that value.  

Also shown in table 3 and fig. 2 is the theoretical prediction $F_L^p(x,Q^2)$ 
for this fit. 
As in the previous fit, and as in \flong, it is calculated using the 
NLO--in--$\alpha_s$ parton
distributions obtained from the fit to $F^m_2(x,Q^2)$ and the order $\alpha_s$
and order $\alpha_s^2$ longitudinal coefficient functions\ref\nlofl{E.B. 
Zijlstra and W. van Neerven, \NP \vyp{B383}{1992}{525}\semi
S.A. Larin and J.A.M. Vermaseren, \ZP {C57}{1993}{93}.}. Hence, I use 
coefficient functions at one higher order for the longitudinal coefficient
function than for $F_2(x,Q^2)$ because they begin at one higher order. 
This is the correct procedure to obtain an expression consistent with the 
use of the two--loop coupling constant in a given renormalization scheme. 
$F_L^p(x,Q^2)$ is a little smaller than that in the first approach in the 
region shown (it is actually larger for $x>0.001$),
with the difference becoming larger at smaller $x$. This qualitative result 
is due to the gluon 
being rather flatter at small $x$ than the gluon in the former approach, 
which grows quickly as $x$ decreases. As one can see from fig. 2 
there is reasonable agreement between the prediction and the measurement,
even though there seems to a systematic trend for the values of 
$F_L^m(x,Q^2)$ to fall as $x$ increases while over most of the range the 
values of $F_L^p(x,Q^2)$ increase slowly with increasing $x$. The 
comparison of $F_L^m(x,Q^2)$ to $F_L^p(x,Q^2)$ yields a $\chi^2$ of 
$13.9$ for $15$ points. Thus, this consistency check gives no real reason 
to distrust the NLO--in--$\alpha_s$ approach. However, it may be argued 
that both the quality of the global fit and the comparison of $F_L^m(x,Q^2)$
with $F_L^p(x,Q^2)$ in the high $y$ region do not give overwhelming 
support for this approach.   

\medskip

\noindent 3. I now consider a fit to precisely the same data but using a 
different 
theoretical framework. I use the LORSC calculation of structure functions 
discussed in \mypaper, which includes all term which are of lowest order in 
$\alpha_s$ for all different types of term in $x$, and does so in terms 
of physical quantities rather than unphysical, 
definition--dependent parton distributions and 
coefficient functions (see \ref\cat{S. Catani, Talk at UK workshop on HERA
physics, September 1995, unpublished; {\tt hep-ph/9609263}, Preprint DDF 
248/4/96, April 1996; Proc. of the International Workshop on Deep Inelastic 
Scattering, Rome, April, 1996, p. 165.} for a definition of physical anomalous 
dimensions). Hence it includes a summation of the so--called
leading $\ln(1/x)$ terms for physical quantities. It is only done to leading 
order because the 
full set of terms required for the NLO calculation is not yet known. The fit
uses leading order $\Lambda_{QCD}^{n_f=4}=100\Mev$, which corresponds to
$\alpha_s(M_Z^2)=0.115$. In this fit the 
normalization required by the H1 data is $1.00$ and for the ZEUS data it is 
$1.015$. The quality of the absolute best fit has a higher confidence level
than that of the NLO--in--$\alpha_s$ fit, i.e. 
$34\%$. Letting this reduce to 
$10\%$ results in a variation of $F_2^p(x,Q^2)$ which again when inserted in 
\four\ leads to negligible effect, and I take the error to be that due to 
$\Delta \tilde \sigma (x,Q^2)$ alone. If I were to let the confidence level 
reduce to $1\%$, as in the NLO--in--$\alpha_s$ fit then the uncertainty
in $F^p_2(x,Q^2)$ would start to increase the error on 
$F^m_L(x,Q^2)$ significantly. 

Using this approach the values of $F_2^p(x,Q^2)$ in the region of $x$
under consideration are about $1.015$ times those in the NLO--in--$\alpha_s$
approach at $Q^2=6\Gev^2$. Taking into account the difference in 
normalization of the data in the two approaches this means that when 
using H1 data the comparison to data is very similar for the two approaches at
this $Q^2$, while for the ZEUS data the LORSC $F^p_2(x,Q^2)$ is slightly 
smaller relative to data than the NLO--in--$\alpha_s$ prediction (due to the 
$2.5\%$ shift in normalization of the data). The growth of $F_2^p(x,Q^2)$ 
with $Q^2$ is greater in the LORSC approach, even taking into account the
normalization difference, and this leads to the better global fit, as 
discussed in \mypaper. This greater growth means that the values of 
$F^m_L(x,Q^2)$ extracted using this approach start out very similar to the 
NLO--in--$\alpha_s$ approach for H1 data and $Q^2$ not much above $6\Gev^2$,
but steadily increase with respect to them as $Q^2$ increases. For the ZEUS 
data the larger normalization shift leads to the LORSC values of $F^p_L(x,Q^2)$
being slightly smaller than those for the NLO--in--$\alpha_s$ fit for 
lowish $Q^2$, but they increase steadily above 
those for the NLO--in--$\alpha_s$ fit at higher $Q^2$. The 
values of $F_L^m(x,Q^2)$  are shown in table 4 and \fig\figthree{Result of 
the measured values of $F^m_L(x,Q^2)$ for the 15 points satisfying the 
cut of 0.3 on the error, along with the predicted values, $F^p_L(x,Q^2)$,
for a LORSC global fit.} and by comparing to table 3 and fig. 2
the trend described above can be easily observed. It 
results in the average value of $F^m_L(x,Q^2)$ being slightly higher in 
this approach than the previous one, and in the tendency of the values of 
$F^m_L(x,Q^2)$ to fall with increasing $x$ or $Q^2$ to reduce. 
The discrepancy between the extracted values for the two theoretical 
approaches, though not large for the points considered in this paper, 
would continue to increase for increasing $Q^2$.  

Also shown in table 4 and fig. 3 is the theoretical prediction $F_L^p(x,Q^2)$ 
for this fit. As discussed in \mypaper\ it is smaller than that in the 
NLO--in--$\alpha_s$ approach in the 
region shown but the difference becomes less at smaller $x$, where the 
NLO--in--$\alpha_s$ curve at constant $y=0.7$ falls as $x$ goes to $0.0001$ 
and below, while in the LORSC approach it stays more constant. 
As one can see from fig. 3 
there is again reasonable agreement between the prediction and the measurement,
though some of the values of $F_L^m(x,Q^2)$ at smaller $x$ lie quite a long 
way above the $F_L^p(x,Q^2)$ curve. The 
comparison of $F_L^m(x,Q^2)$ to $F_L^p(x,Q^2)$ yields a $\chi^2$ of 
$12.4$ for $15$ points. This is very slightly better than in the 
NLO--in--$\alpha_s$ approach. Hence, this consistency check is certainly 
satisfactory, giving no evidence for the failure of the LORSC approach.
Indeed, it is even slightly better than the NLO--in--$\alpha_s$ approach
which has a rather less successful global fit to data. 

\medskip

\noindent 4. Finally I consider the consequences of using a simple 
LO--in--$\alpha_s$ calculation of the structure functions. Methods of 
fitting the small $x$ data based simply on the LO--in--$\alpha_s$ 
calculations have not been uncommon in recent years, the ``double asymptotic
scaling'' formula comes from this calculation \ref\das{R.D. Ball and S. Forte, 
\PL \vyp{335}{1994}{77}.}, and have been 
advertised as being relatively successful, especially if $Q^2$ is not too 
low. I find that the standard approach is rather too simplistic, 
and in fact the LO--in--$\alpha_s$ approach is completely ruled out. 
If one performs a global fit to structure function data other than 
that used in the extraction of $F_L^m(x,Q^2)$ and using simply 
those values of $F_2(x,Q^2)$ advertised in \hone\ and \zeus, then the 
result, requiring the same normalizations as the NLO--in--$\alpha_s$ 
approach, is very nearly as good as for the NLO--in--$\alpha_s$ fit
over the whole $x$ range. 
It yields a $\chi^2$ of $\sim 20$ more for $\sim 1100$ points and a 
confidence level of $2\%$, compared to the earlier $6\%$. However, the 
quality of the fit to the small $x$ data is achieved at the expense of having
a fairly large coupling constant, one--loop $\alpha_s(M_z^2)=0.124$ and a much 
larger small $x$ gluon than the NLO--in--$\alpha_s$. This,
coupled with the fact the the NLO corrections to $F_L(x,Q^2)$ are negative 
at small $x$, leads to $F^p_L(x,Q^2)$ being a great deal larger at small 
$x$ than in the NLO--in--$\alpha_s$ approach. 
Moreover, as in the NLO--in--$\alpha_s$ 
approach it is the small $x$ -- large $y$ region where $F^p_2(x,Q^2)$ is 
beginning to systematically undershoot data. Hence, correcting the values of 
$F_2^m(x,Q^2)$ in the fit above for the correct values of $F_L^p(x,Q^2)$
leads to a worsening of the fit by $\sim 25$ (even after the fit is redone)
and a consequent reduction of 
the confidence level to $0.5\%$. This effect, which is always ignored in
LO--in--$\alpha_s$ fits to structure function data, clearly weakens the 
support for this type of approach considerably.     

It is when I perform the consistency check between the values of 
$F^m_L(x,Q^2)$ and $F^p_L(x,Q^2)$, i.e. examine those points which are most 
sensitive to $F_L(x,Q^2)$, that the most dramatic effects are seen. 
The values of $F_2^p(x,Q^2)$ are fairly similar to those in the 
NLO--in--$\alpha_s$ case, and hence so are the values of $F^m_L(x,Q^2)$. Also,
any significant change in the extrapolated $F_2^p(x,Q^2)$ reduces the 
confidence level of the fit to unacceptable levels, so once again the error 
on $F_L^m(x,Q^2)$ is taken to be that on $\tilde \sigma (x,Q^2)$ alone. 
However, the values of $F_L^p(x,Q^2)$ are very much larger than in the 
previous approaches. The set of values of $F^m_L(x,Q^2)$ and $F_L^p(x,Q^2)$ 
are shown in table 5 and displayed in \fig\figfour{Result of 
the measured values of $F^m_L(x,Q^2)$ for the 15 points satisfying the 
cut of 0.3 on the error, along with the predicted values, $F^p_L(x,Q^2)$,
for a LO--in--$\alpha_s$ global fit.} . It is immediately clear that
the measured values undershoot the predicted values significantly.  
A calculation of the $\chi^2$ is not necessary to prove that an 
extrapolation into the region where both $F_2(x,Q^2)$ and $F_L(x,Q^2)$ 
make important contributions to the cross--section rules out a
LO--in--$\alpha_s$ calculation of structure functions completely.  

An alternative way to demonstrate this is to compare the values of 
$F_2^p(x,Q^2)$
obtained from the fit to the low $y$ data to the ``measured'' values 
$F^m_2(x,Q^2)$ obtained from the measurement of the cross--section and the 
predicted values of $F_L^p(x,Q^2)$. This is shown for two $x$ bins 
in \fig\figfive{The ``measured'' values of the structure function, 
$F^m_2(x,Q^2)$ (data points), compared to the theoretical values, 
$F_2^p(x,Q^2)$,
obtained from the best fit (solid line) in two different small $x$ bins,
when using the LO--in--$\alpha_s$ calculation.},
and it is very easy to see that the large predicted $F^p_L(x,Q^2)$ 
combined with the measured $\tilde \sigma (x,Q^2)$ leads to a large
upturn in the data, above the theoretical curve, as $y$ becomes larger. A 
very similar effect is seen in all the different $x$ bins, and is clear 
evidence that a LO--in--$\alpha_s$ fit is completely ruled out by high $y$
data. I note that in this LO--in--$\alpha_s$ fit a large component of the 
small $x$ $F_2(x,Q^2)$ comes from the steepness of the input quark. If I 
were to follow the reasoning behind double asymptotic scaling, i.e. that 
the small--$x$ inputs should be flat at low $Q^2$, and the rise generated by 
evolution, then not only would the quality of the global fit fall, but
the value of $F^p_L(x,Q^2)$ would be even larger, and this clear discrepancy
between theory and data at high $y$ would become even more pronounced. 
Therefore, this approach is very strongly ruled out in the region of 
small $x$ and large $y$. The fact
that the LO--in--$\alpha_s$ approach fails badly, but acquires very 
important small--$x$ corrections at NLO--in--$\alpha_s$ also implies that 
the large small--$x$ corrections at higher orders in $\alpha_s$ should be
important, although, as already mentioned, there is no overwhelming 
evidence for this yet.
 
\medskip

The three different theoretical approaches used above are far from being the 
full set proposed to describe small $x$ physics, and this 
technique can be applied to others. However, perhaps rather alarmingly they
are the only approaches which have been used in fits to a global range
of data and which are fully constrained (except for approximate fits
using factorization--scheme--dependent, and hence incorrect methods of 
including leading $\ln(1/x)$ terms within the collinear factorization 
framework), and considerable work would be required to use any other 
approach. 

I also note that at rather higher $x$ than considered in this 
paper there have been a number of direct measurements of $F_L(x,Q^2)$:
the SLAC hydrogen and deuterium scattering experiments \ref\SLACh{L.W. Whitlow
{\it et al.}, \PL \vyp{B250}{1990}{193.}}, the SLAC E140
experiments \ref\SLACe{E140 Collaboration: S. Dasu {\it et al.},
\PRL \vyp{61}{1988}{1061}\semi E140 Collaboration: S. Dasu {\it et al.},
\PR \vyp{D49}{1994}{5641}\semi E140X Collaboration: L.H. Tao {\it et al.},
\ZP \vyp{C70}{1996}{387}.}, CDHSW \ref\CDHSW{CDHSW Collaboration: P. Berge 
{\it et al.}, \ZP \vyp{C49}{1991}{187}.}, BCDMS \ref\BCDMSfl{BCDMS 
Collaboration: A.C. Benvenuti {\it et al.},
\PL \vyp{B195}{1987}{91}.}, CCFR \ref\ccfrfl{A. Bodek for the CCFR/NuTeV
collaboration; Proc. of the International Workshop on Deep Inelastic 
Scattering, Rome, April, 1996, p. 213.} and NMC \nmc. This data has 
fairly large error bars, and much is in the region of the charm threshold.
However, it seems that with a better treatment of this threshold than 
used in this paper then either 
the NLO--in--$\alpha_s$ approach or the LORSC approach would fit the data 
fairly well, with the former tending to be perhaps a little high, 
and the latter 
perhaps a little low. (The $R_QCD$ curve in \nmc\ should be treated with 
caution since it uses the gluon from \hone, which has been criticised
above, and a LO--in--$\alpha_s$ formula using four massless
quarks rather than the appropriate NLO--in--$\alpha_s$ formula with 
massive charm quarks.) The very low $Q^2$ data from these 
measurements gives some evidence for higher 
twist effects \ref\bodek{A. Bodek, S. Rock and U.K. Yang, \ZP to appear.}.   
 
I finally note that during the preparation of this article some 
new preliminary data at high $y$ has been released by H1 \ref\newdata{V.
Chekelian, plenary talk ``The Structure of Hadrons'', Lepton-Photon Symposium,
Hamburg, 1996.}. There is new data at $y=0.7$ for the same values of
$Q^2$, except for the lowest $Q^2$ bin, which seems to improve the comparison
for both the NLO--in--$\alpha_s$ fit and the LORSC fit, i.e. the measured
cross--section for $Q^2=12\Gev^2$ and $20\Gev^2$ increase, while that 
for $25\Gev^2$ goes down a little. The improvement for the LORSC fit appears
to be better than that for the NLO--in--$\alpha_s$ fit. The comparison for 
the LO--in--$\alpha_s$ will get even worse. There is also some data at 
$y=0.82$ with large error bars. Here the cross--section measurements look
rather low. The QCD study is presumably performed 
in the same manner as in \flong. A detailed study in the manner of this 
paper will await final data.  

\newsec{Conclusions.}

In this paper I have examined the implications of the idea proposed in 
\flong\ of extrapolating a fit performed to structure function data at 
relatively low $y$ into the region of high $y$, where the longitudinal 
structure function starts to make an impact on the cross--section. In 
\flong\ it was assumed that believing the extrapolation of 
$F_2(x,Q^2)$ into this high $y$ region using a given theoretical framework, 
the difference between the extrapolation and the measured cross--section
would give a measurement of $F_2(x,Q^2)$. In this paper I have worked on a 
different principle, pointing out that different theoretical models will 
lead to different extrapolations, and also different predictions for 
$F_L(x,Q^2)$. All one is really examining is whether, in the region 
where both structure functions contribute to the cross-section, the 
data on the cross--section are consistent with the theory. One way to 
present this is to compare the ``measured'' values of $F_L(x,Q^2)$ with the 
predicted values. 

I have first done this using the same theoretical framework as in \flong\
(up to a different treatment of charm mass effects) and fitting to the same 
``low $y$'' data. This provides results almost identical to \flong,
which may be naively interpreted as consistency of the NLO--in--$\alpha_s$ 
structure functions at high $y$. However, I have also argued very strongly 
that this type of fit is very badly underconstrained, particularly 
concerning the gluon for $x>0.1$, and is not useful. 
I have then repeated the procedure using a NLO--in--$\alpha_s$ global fit  
at low $y$ 
which is a great deal more constraining. I have also used points in the latter
of \zeus, and also in \hone, which give similar uncertainty on an 
extracted $F_L(x,Q^2)$ to those points in \flong. Even using this same 
theoretical framework, the conclusion concerning the measured values of 
$F_L(x,Q^2)$ changes somewhat, with the average value being $\sim 60\%$ 
those in \flong. The agreement between the ``measured'' and predicted 
values is good. 

Using two further theoretical approaches, the LORSC
calculation and the LO--in--$\alpha_s$ calculation fits to the same 
low $y$ data the ``measured'' values of $F_L(x,Q^2)$ do not change by
very large amounts, although the changes would quickly increase with 
increasing $y$. The predictions for $F_L(x,Q^2)$ are however different
to the previous case: for the LORSC approach consistency is in fact
very slightly better than the NLO--in--$\alpha_s$ approach, but in the 
LO--in--$\alpha_s$ approach the predicted $F_L(x,Q^2)$ is far too large.
Thus, the LO--in--$\alpha_s$ approach, and consequently (and
particularly) double asymptotic scaling, is ruled out precisely in the 
region where it is claimed most strongly to hold, i.e. high $Q^2$ and low $x$. 

Hence, within this limited study I conclude that both the 
NLO--in--$\alpha_s$ and the LORSC calculations are consistent with high $y$ 
cross--section data. Therefore, forgetting other possible theoretical models,
the values of $F_L(x,Q^2)$ are very likely to be similar to the predictions 
of these approaches when they are constrained by a global fit to 
structure function data. By definition this means that the values are 
similar to those ``measured'' using the two approaches, but somewhat lower 
than those presented in \flong. An increase in the precision of measurement 
of the high $y$ cross-section, or an extension to slightly higher values of
$y$ would be very important in differentiating between the two theoretical
approaches, and potentially any others.  

Alternatively, a direct measurement of $F_L(x,Q^2)$ would be even more 
useful. In this case there would not be any inbuilt uncertainty due to 
a particular (and always to some degree approximate) theoretical model, 
or to the measurement depending on 
how accurately, or correctly, unknown parameters are determined by the 
fit to low $y$ data (even though this is probably small). It would also 
eliminate the significant
possibility that incorrectness in both $F_2(x,Q^2)$ and
$F_L(x,Q^2)$ in a given theoretical approach could act to partially,
or even largely cancel out and still lead to apparent consistency in the above 
approach. Hence I
encourage strongly any attempt to measure $F_L(x,Q^2)$ directly using
any method as a way to help discriminate strongly between 
different theoretical approaches to calculating small $x$ structure 
functions, and to obtain real data on a real physical quantity. 

\vfill 
\eject

\medskip
\noindent{\bf Acknowledgements.}
\medskip
I would like to thank R.G. Roberts for continual help during the period of 
this work and for the use of the MRS fit program. I would also like to
thank Mandy Cooper--Sarkar, Robin Devenish and Norman McCubbin for
useful discussions. 

\vfill 
\eject

\noindent {\bf Table 1}\hfil\break
\noindent Data points with sufficiently high $y$ and low enough error on 
$\tilde \sigma$ for the error on $F^m_L(x,Q^2)$ to be less than 
$0.3$. The $Q^2$ values are denoted in $\Gev^2$.

\medskip

\hfil\vtop{{\offinterlineskip
\halign{ \strut\tabskip=0.6pc
\vrule#&  \hfil#&  \vrule#&  \hfil#& \vrule#& \hfil#& \vrule#& \hfil#&
\vrule#& \hfil#& \vrule#&  \hfil#&  \vrule#&  \hfil#& \vrule#& \hfil#& 
\vrule#& \hfil#& \vrule#& \hfil#& \vrule#\tabskip=0pt\cr
\noalign{\hrule}
& \omit &\omit& \multispan{5}\hfil\bf H1 data \hfil &\omit& \omit &&
\omit &\omit& \multispan{5}\hfil\bf ZEUS data \hfil &\omit& \omit &\cr
\noalign{\hrule}
& $Q^2$ && $x\cdot10^4$ && $y$ &&$\tilde \sigma$ &&$\Delta \tilde \sigma$ &&
$Q^2$ && $x\cdot10^4$ && $y$ &&$\tilde \sigma$ &&$\Delta \tilde \sigma$ &\cr
\noalign{\hrule}
& 8.5  && 1.35 && 0.70 && 1.165 && 0.099 && 6.5  && 1.00 && 0.72 && 
1.092 && 0.084 &\cr
& 12.0 && 1.90 && 0.70 && 1.198 && 0.079 && 8.5  && 1.60 && 0.59 
&& 1.248 && 0.074 &\cr
& 15.0 && 2.38 && 0.70 && 1.368 && 0.085 && 10.0 && 1.60 && 0.69 
&& 1.225 && 0.114 &\cr
& 20.0 && 3.17 && 0.70 && 1.276 && 0.079 && 12.0 && 2.50 && 0.53 
&& 1.233 && 0.068 &\cr
& 25.0 && 3.96 && 0.70 && 1.439 && 0.089 && 15.0 && 2.50 && 0.67 
&& 1.424 && 0.108 &\cr
& 25.0 && 5.00 && 0.56 && 1.430 && 0.069 && 27.0 && 6.30 && 0.48 
&& 1.407 && 0.053 &\cr
& 35.0 && 5.54 && 0.70 && 1.435 && 0.099 && 35.0 && 6.30 && 0.62 
&& 1.459 && 0.086 &\cr
& \omit && \omit && \omit && \omit && \omit && 60.0 && 10.0 
&& 0.67 && 1.438 && 0.088 &\cr
\noalign{\hrule}}}}\hfil

\vskip 0.3in

\noindent {\bf Table 2}{\hfil\break
\noindent The ``measurement'' $F^m_L(x,Q^2)$
and prediction $F^p_L(x,Q^2)$ of $F_L(x,Q^2)$ for the 
NLO--in--$\alpha_s$ fit performed in a manner analogous to that in \flong,
and for the data in \flong\ only. Also shown are the ``measured'' values 
$F^m_L(x,Q^2)[4]$ in \flong.

\medskip

\hfil\vtop{{\offinterlineskip
\halign{ \strut\tabskip=0.6pc
\vrule#&  \hfil#&  \vrule#&  \hfil#& \vrule#& \hfil#& \vrule#&
\hfil#& \vrule#& \hfil#& \vrule#& \hfil#& \vrule#\tabskip=0pt\cr
\noalign{\hrule}
& \omit &\omit& \multispan{7}\hfil\bf H1 data \hfil &\omit& \omit &\cr
\noalign{\hrule}
& $Q^2$ && $x\cdot10^4$ && $F^p_L$ && $F_L^m$ && $\Delta F^m_L$ 
&& $F_L^m[4]$ &\cr
\noalign{\hrule}
& 8.5  && 1.35 && 0.31 && 0.51 && 0.22 && 0.51 &\cr
& 12.0 && 1.90 && 0.34 && 0.62 && 0.18 && 0.63 &\cr
& 15.0 && 2.38 && 0.35 && 0.33 && 0.19 && 0.35 &\cr
& 20.0 && 3.17 && 0.36 && 0.61 && 0.17 && 0.67 &\cr
& 25.0 && 3.96 && 0.39 && 0.30 && 0.20 && 0.33 &\cr
& 35.0 && 5.54 && 0.38 && 0.33 && 0.22 && 0.39 &\cr
\noalign{\hrule}}}}\hfil

\vskip 0.3in

\vfill
\eject

\noindent {\bf Table 3}\hfil\break
\noindent $F^m_L(x,Q^2)$ and $F^p_L(x,Q^2)$ for the NLO--in--$\alpha_s$ 
global fit.

\medskip

\hfil\vtop{{\offinterlineskip
\halign{ \strut\tabskip=0.6pc
\vrule#&  \hfil#&  \vrule#&  \hfil#& \vrule#& \hfil#& \vrule#& \hfil#&
\vrule#& \hfil#& \vrule#&  \hfil#&  \vrule#&  \hfil#& \vrule#& \hfil#& 
\vrule#& \hfil#& \vrule#& \hfil#& \vrule#\tabskip=0pt\cr
\noalign{\hrule}
& \omit &\omit& \multispan{5}\hfil\bf H1 data \hfil &\omit& \omit &&
\omit &\omit& \multispan{5}\hfil\bf ZEUS data \hfil &\omit& \omit &\cr
\noalign{\hrule}
& $Q^2$ && $x\cdot10^4$ && $F^p_L$ && $F_L^m$ &&$\Delta F^m_L$ &&
$Q^2$ && $x\cdot10^4$ && $F^p_L$ && $F_L^m$ &&$\Delta F^m_L$ &\cr
\noalign{\hrule}
& 8.5  && 1.35 && 0.24 && 0.41 && 0.22 && 6.5 && 1.00 && 0.19 
&& 0.39 && 0.17 &\cr
& 12.0 && 1.90 && 0.28 && 0.51 && 0.18 && 8.5  && 1.60 && 0.23
&& 0.16 && 0.25 &\cr
& 15.0 && 2.38 && 0.30 && 0.21 && 0.19 && 10.0 && 1.60 && 0.26 
&& 0.34 && 0.26 &\cr
& 20.0 && 3.17 && 0.31 && 0.50 && 0.17 && 12.0 && 2.50 && 0.270 
&& 0.42 && 0.29 &\cr
& 25.0 && 3.96 && 0.35 && 0.19 && 0.20 && 15.0 && 2.50 && 0.29
&& 0.03 && 0.27 &\cr
& 25.0 && 5.00 && 0.33 && 0.02 && 0.27 && 27.0 && 6.30 && 0.32
&&-0.18 && 0.30 &\cr
& 35.0 && 5.54 && 0.38 && 0.22 && 0.22 && 35.0 && 6.30 && 0.33
&& 0.05 && 0.26 &\cr
& \omit && \omit && \omit && \omit && \omit && 60.0 && 10.0 
&& 0.32 && 0.07 && 0.22 &\cr
\noalign{\hrule}}}}\hfil

\vskip 0.3in

\noindent {\bf Table 4}\hfil\break
\noindent $F^m_L(x,Q^2)$ and $F^p_L(x,Q^2)$ for the LORSC 
global fit.

\medskip

\hfil\vtop{{\offinterlineskip
\halign{ \strut\tabskip=0.6pc
\vrule#&  \hfil#&  \vrule#&  \hfil#& \vrule#& \hfil#& \vrule#& \hfil#&
\vrule#& \hfil#& \vrule#&  \hfil#&  \vrule#&  \hfil#& \vrule#& \hfil#& 
\vrule#& \hfil#& \vrule#& \hfil#& \vrule#\tabskip=0pt\cr
\noalign{\hrule}
& \omit &\omit& \multispan{5}\hfil\bf H1 data \hfil &\omit& \omit &&
\omit &\omit& \multispan{5}\hfil\bf ZEUS data \hfil &\omit& \omit &\cr
\noalign{\hrule}
& $Q^2$ && $x\cdot10^4$ && $F^p_L$ && $F_L^m$ &&$\Delta F^m_L$ &&
$Q^2$ && $x\cdot10^4$ && $F^p_L$ && $F_L^m$ &&$\Delta F^m_L$ &\cr
\noalign{\hrule}
& 8.5  && 1.35 && 0.20 && 0.43 && 0.22 && 6.5 && 1.00 && 0.19 
&& 0.37 && 0.17 &\cr
& 12.0 && 1.90 && 0.22 && 0.53 && 0.18 && 8.5  && 1.60 && 0.19
&& 0.12 && 0.25 &\cr
& 15.0 && 2.38 && 0.22 && 0.25 && 0.19 && 10.0 && 1.60 && 0.21 
&& 0.33 && 0.26 &\cr
& 20.0 && 3.17 && 0.22 && 0.55 && 0.17 && 12.0 && 2.50 && 0.20 
&& 0.37 && 0.29 &\cr
& 25.0 && 3.96 && 0.24 && 0.24 && 0.20 && 15.0 && 2.50 && 0.22
&& 0.03 && 0.27 &\cr
& 25.0 && 5.00 && 0.23 && 0.09 && 0.27 && 27.0 && 6.30 && 0.21
&&-0.18 && 0.30 &\cr
& 35.0 && 5.54 && 0.23 && 0.30 && 0.22 && 35.0 && 6.30 && 0.22
&& 0.09 && 0.26 &\cr
& \omit && \omit && \omit && \omit && \omit && 60.0 && 10.0 
&& 0.21 && 0.13 && 0.22 &\cr
\noalign{\hrule}}}}\hfil

\vskip 0.3in

\vfill
\eject

\noindent {\bf Table 5}\hfil\break
\noindent $F^m_L(x,Q^2)$ and $F^p_L(x,Q^2)$ for the LO--in--$\alpha_s$ 
global fit.

\medskip

\hfil\vtop{{\offinterlineskip
\halign{ \strut\tabskip=0.6pc
\vrule#&  \hfil#&  \vrule#&  \hfil#& \vrule#& \hfil#& \vrule#& \hfil#&
\vrule#& \hfil#& \vrule#&  \hfil#&  \vrule#&  \hfil#& \vrule#& \hfil#& 
\vrule#& \hfil#& \vrule#& \hfil#& \vrule#\tabskip=0pt\cr
\noalign{\hrule}
& \omit &\omit& \multispan{5}\hfil\bf H1 data \hfil &\omit& \omit &&
\omit &\omit& \multispan{5}\hfil\bf ZEUS data \hfil &\omit& \omit &\cr
\noalign{\hrule}
& $Q^2$ && $x\cdot10^4$ && $F^p_L$ && $F_L^m$ &&$\Delta F^m_L$ &&
$Q^2$ && $x\cdot10^4$ && $F^p_L$ && $F_L^m$ &&$\Delta F^m_L$ &\cr
\noalign{\hrule}
& 8.5  && 1.35 && 0.61 && 0.35 && 0.22 && 6.5 && 1.00 && 0.60 
&& 0.33 && 0.17 &\cr
& 12.0 && 1.90 && 0.60 && 0.45 && 0.18 && 8.5  && 1.60 && 0.58
&& 0.07 && 0.25 &\cr
& 15.0 && 2.38 && 0.59 && 0.17 && 0.19 && 10.0 && 1.60 && 0.61 
&& 0.28 && 0.26 &\cr
& 20.0 && 3.17 && 0.57 && 0.47 && 0.17 && 12.0 && 2.50 && 0.56 
&& 0.33 && 0.29 &\cr
& 25.0 && 3.96 && 0.60 && 0.17 && 0.20 && 15.0 && 2.50 && 0.58
&& -0.01 && 0.27 &\cr
& 25.0 && 5.00 && 0.56 && 0.01 && 0.27 && 27.0 && 6.30 && 0.51
&&-0.18 && 0.30 &\cr
& 35.0 && 5.54 && 0.55 && 0.23 && 0.22 && 35.0 && 6.30 && 0.53
&& 0.07 && 0.26 &\cr
& \omit && \omit && \omit && \omit && \omit && 60.0 && 10.0 
&& 0.45 && 0.13 && 0.22 &\cr
\noalign{\hrule}}}}\hfil

\footatend\vfill\supereject\immediate\closeout\rfile\writestoppt
\baselineskip=14pt\centerline{{\bf References}}\bigskip{\frenchspacing%
\parindent=20pt\escapechar=` \input refs.tmp\vfill\eject}\nonfrenchspacing

\vfill\eject\immediate\closeout\ffile{\parindent40pt
\baselineskip14pt\centerline{{\bf Figure Captions}}\nobreak\medskip
\escapechar=` \input figs.tmp\vfill\eject}

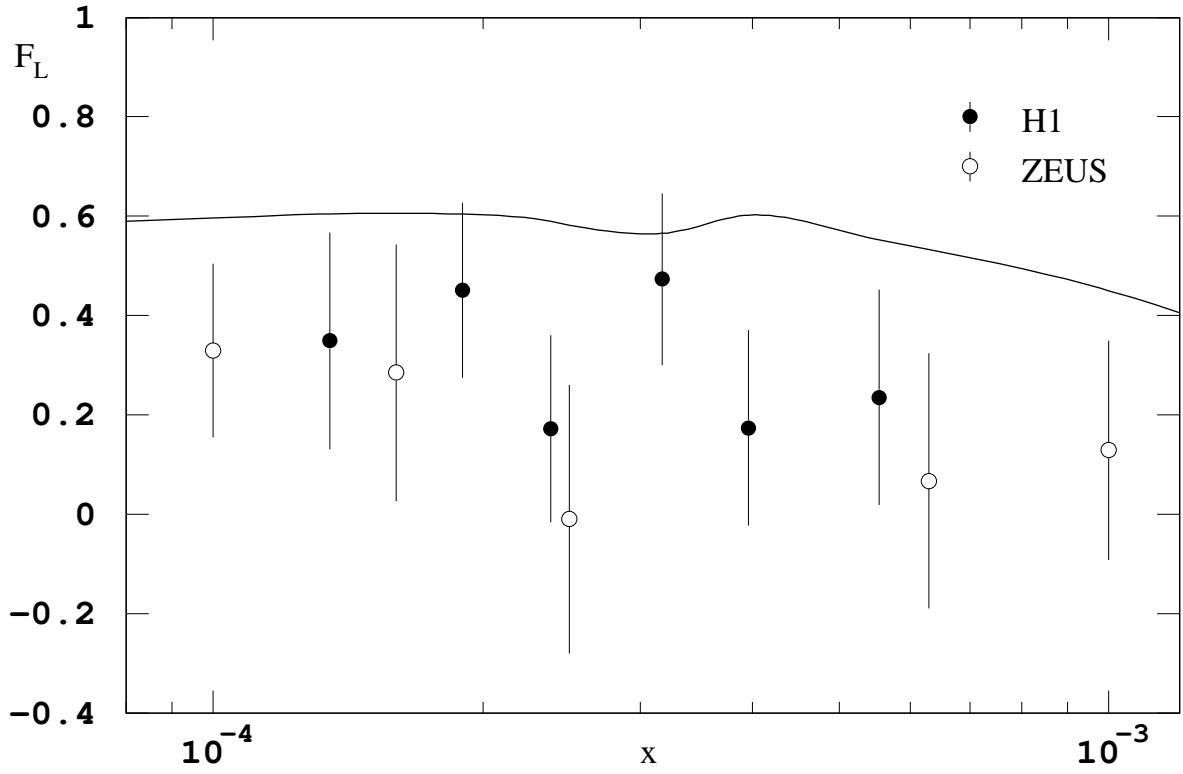
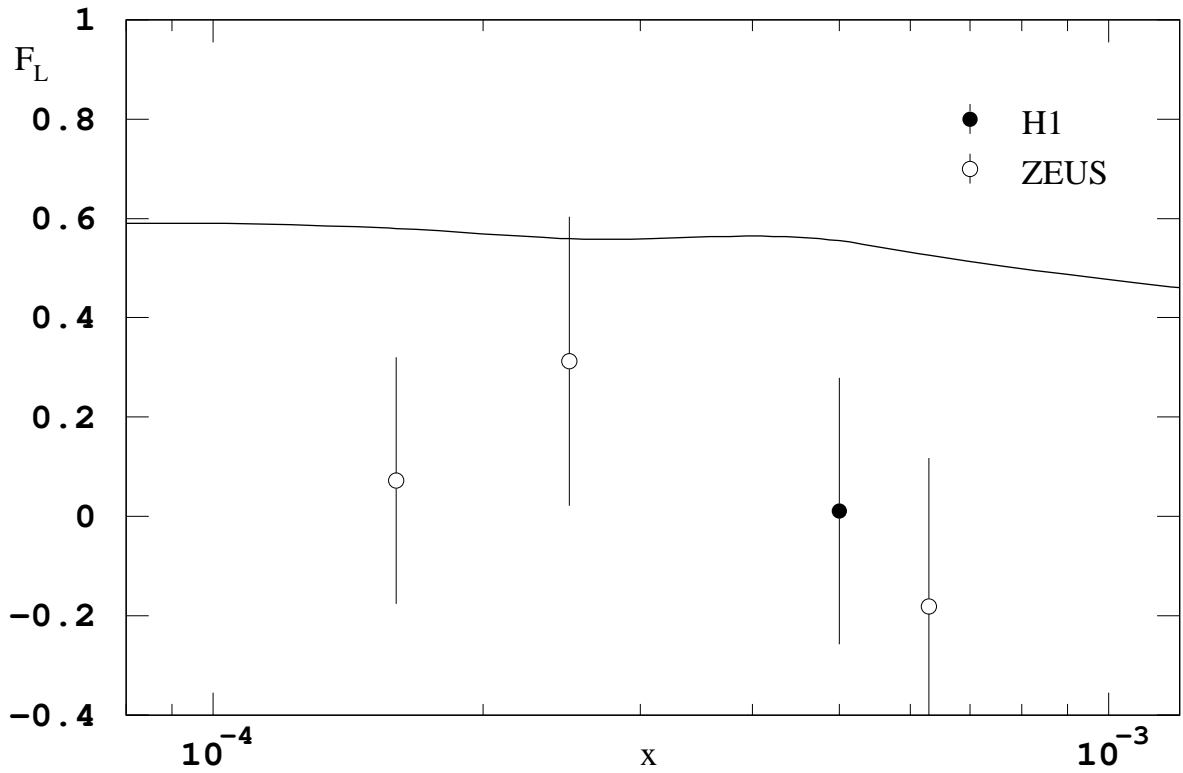
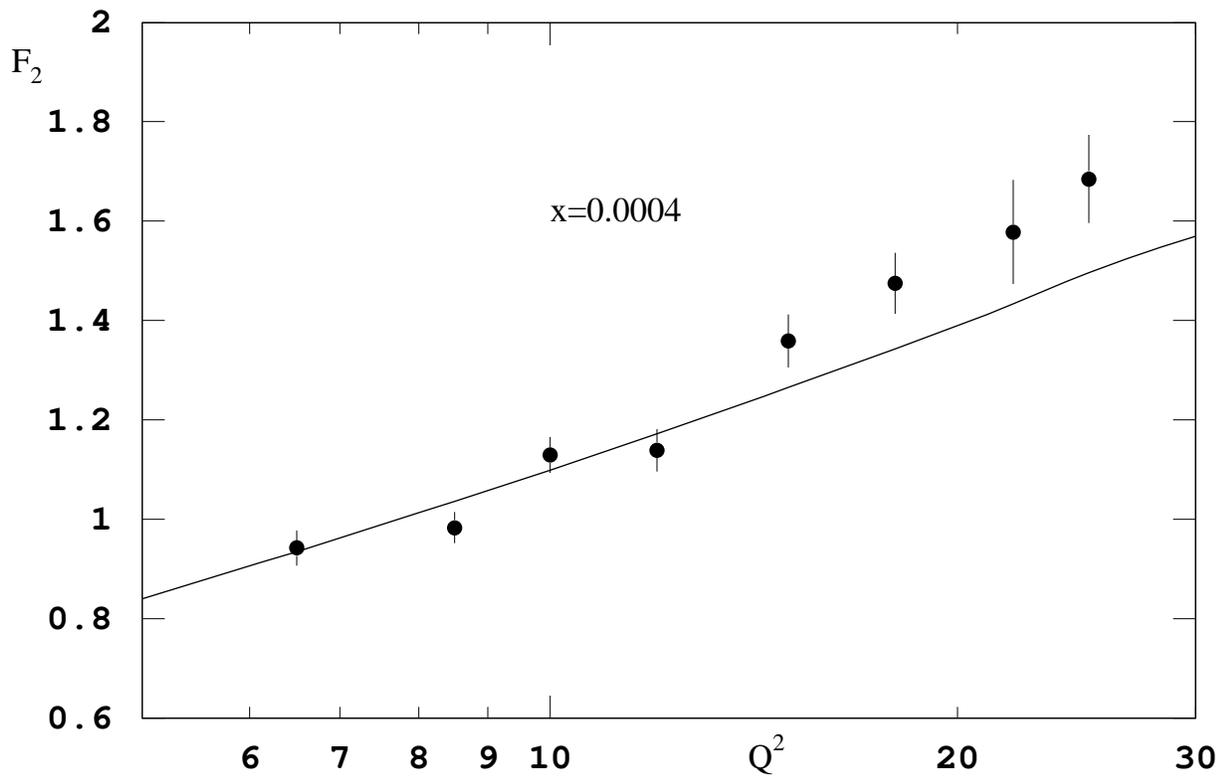
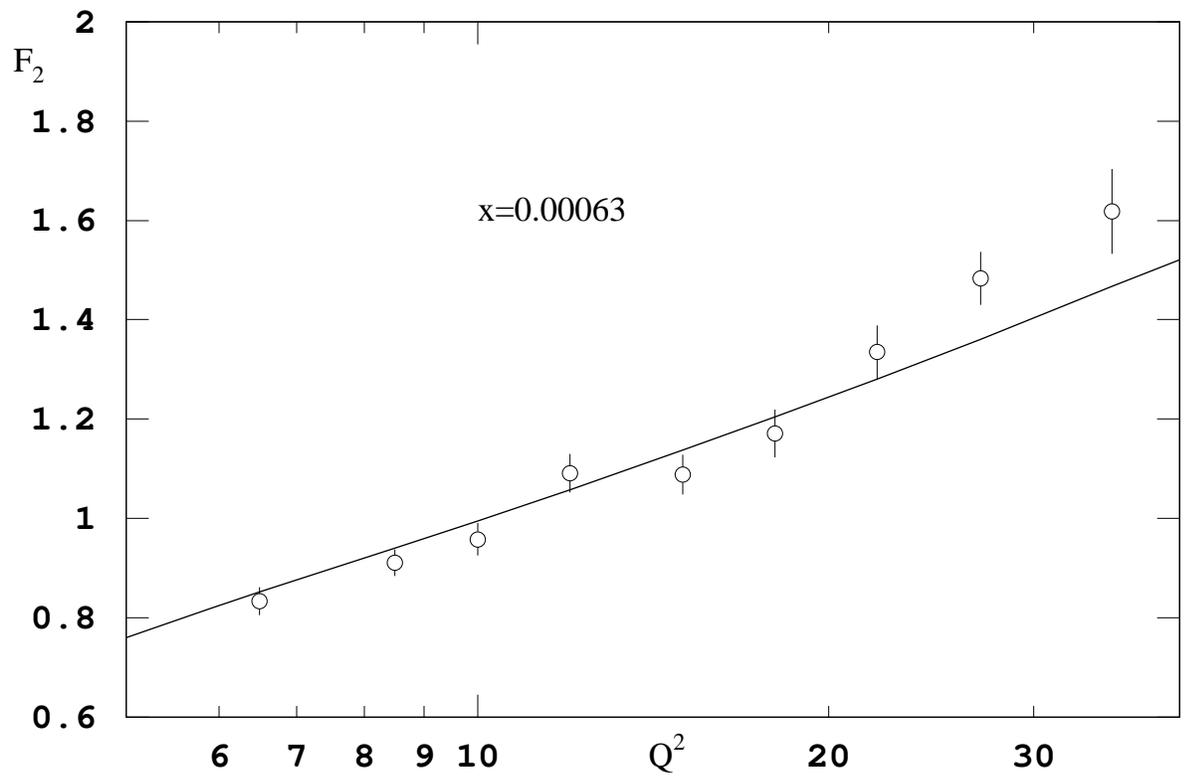
\end